\documentclass[%
 reprint,
showpacs,preprintnumbers,
 amsmath,amssymb,
 aps,
]{revtex4-1}

\usepackage{graphicx}% Include figure files
\usepackage{dcolumn}% Align table columns on decimal point
\usepackage{bm}% bold math
\usepackage{float}
\usepackage{color}

\begin{document}

\title{Indirect exchange interaction between magnetic impurities in one-dimensional gapped helical states}% Force line breaks with \\
\author{Zahra Karimi}
\author{Mir Vahid Hosseini}
 \email[Corresponding author: ]{mv.hosseini@znu.ac.ir}
\author{Jamal Davoodi}
\affiliation{Department of Physics, Faculty of Science, University of Zanjan, Zanjan 45371-38791, Iran}
\date{\today}

\begin{abstract}
We investigate theoretically indirect exchange interaction between magnetic impurities mediated by one-dimensional gapped helical states. Such states, containing massive Dirac fermions, may be realized on the edge of a two-dimensional topological insulator when time-reversal symmetry is weakly broken. We find that the indirect exchange interaction consists of Heisenberg, Dzyaloshinsky-Moriya, in-plane and out-of-plane Ising terms. These terms decay exponentially when Fermi level lies inside the bandgap whereas the Dzyaloshinsky-Moriya term has smallest amplitude. Outside the energy gap the massive helical states modify oscillatory behaviors of the range functions so that their periods decrease near the edge of band in terms of energy gap or Fermi energy. In addition, the out-of-plane Ising term vanishes in the case of zero-gap structure but its oscillation amplitude increases versus energy gap and decreases as a function of Fermi energy whereas the oscillation amplitudes of other components remain constant. Analytical results are also obtained for subgap and over gap regimes. Furthermore, the effects of electron-electron interactions are analyzed.
\end{abstract}

%\pacs{75.10.-b, 75.30.Hx, 75.70.Tj, 75.75.-c}
\maketitle
%%%%%%%%%%%%%%%%%%%%%%%%%%%%%%%%%%%%%%%%%%%%%%%%%%%%%%%%%%%%%%%%%%%%%%%%%%%
\section {Introduction} \label{s1}
%%%%%%%%%%%%%%%%%%%%%%%%%%%%%%%%%%%%%%%%%%%%%%%%%%%%%%%%%%%%%%%%%%%%%%%%%%%
Quantum nature of phenomena becomes more pronounced in low dimensional systems \cite{Giamarchi}. Especially, in recent years much attention in condensed matter physics has been paid to two-dimensional systems \cite{2Dcrystal} due to synthesis of graphene \cite{Graphene} and related materials \cite{2DDirac} with Dirac-like dispersion relation. Materials with linearly dispersing spectrum have attracted considerable amount of attention due to providing new opportunities for both fundamental aspects and potential applications.
Moreover, the existence of Weyl fermions has been predicted \cite{WSMRev1} and reported \cite{WSMRev2} in Weyl semimetals that are three-dimensional analogs of graphene. However, one-dimensional (1D) Dirac materials provide a promising alternative to studying exotic characteristics of chiral quantum states. It may raise an interesting perspective in a variety of contexts. In particular, since features of indirect exchange interactions between magnetic impurities depend on both the dimensionality and the band dispersion of host materials, one may expect a unique behavior in 1D Dirac materials.

The study of indirect exchange interactions between magnetic impurities in one-, two- and three-dimensional materials with parabolic band structure goes back to the seminal works by Ruderman, Kittel, Kasuya, and Yosida (RKKY) \cite{RKKYPertu1,RKKYPertu2,RKKYPertu3}. This interaction, known also as RKKY interaction, has been investigated in new systems whose band structures disperse linearly with chiral feature, for instance, graphene \cite{RKKYGraphene1,RKKYGraphene2,RKKYGraphene3} and phosphorene \cite{RKKYphosphorene1,RKKYphosphorene2} in two dimensions and  Weyl/Dirac semimetals \cite{RKKYWeyl1,RKKYWeyl2,RKKYWeyl3,RKKYWeyl4} in three dimensions. Moreover, exotic spin textures due to spin-momentum locking on the surface of three-dimensional \cite{3DTIRKKY0,3DTIRKKY1,3DTIRKKY2,3DTIRKKY3,3DTIRKKY4,3DTIRKKY5,3DTIRKKYGap} and at the edge of two-dimensional \cite{2DTIRKKY1,2DTIRKKY2,2DTIRKKY3,RKKYeeHelicLiq,RKKYKondeeHelicLiq} topological insulators have been predicted as another example of nontrivial RKKY interaction. Another work has discussed RKKY interaction near the helical edge by taking into account of both bulk and edge modes \cite{2DnearEdge}. Furthermore, it has been shown that magnetic exchange interaction is adjustable by a vertical bias in thick films \cite{RKKYThickThinFilm} or a thin slab of topological insulators \cite{RKKYThickThinFilm,RKKYthinFilm}. Recently, the RKKY interaction mediated by surface states \cite{2DRKKYTS} and helical Majorana edge states \cite{MajoTScRKKY1,MajoTScRKKY2} in a topological superconductor was also investigated.

Several studies have been performed on magnetic topological insulators \cite{DopgapSize1} elucidating magnetic properties of promising topological insulator candidate materials such as HgTe quantum wells \cite{BHZTI}, Bi$_2$Se$_3$ \cite{TIBiSe}, and Sb$_2$Te$_3$ \cite{TISbTe}. Experimentally, magnetic doping of topological insulators with transition metal \cite{DopgapSize1} as well as rare earth \cite{REdop} ions has been studied recently. In the former case impurity magnetic moments are ferromagnetically ordered resulting in massive Dirac Fermions \cite{Masive1,Masive2} and realizing quantum anomalous Hall effect \cite{QAHE}. It has been shown that out-of-plane magnetic ordering can be observed with Cr dopants \cite{Cr-doped1,Cr-doped2,Cr-doped3}. More recently, using synchrotron-based x-ray techniques, it has been demonstrated that the topological insulator surface is magnetically ordered while the bulk is not in Cr-doped Bi$_2$Se$_3$ \cite{surfOrder}. In the case of rare earth doping, although high magnetic moments can be introduced into the topological insulators but, instead of long range ferromagnetic ordering, antiferromagnetic ordering along with Dirac gap opening have been reported for Gd-doped topological insulators \cite{AntiFerRE}. Moreover, a large effective magnetization can be introduced in rare-earth-doped topological insulators by proximity coupling either to a ferromagnetic insulator \cite{FeroHetro} or in a heterostructure with transition-metal-doped layers \cite{TM-REHetro1,TM-REHetro2}.

Theoretically, on the other hand, it has been revealed conductance properties of topological boundary states to be coupled to magnetic impurities \cite{CondTITheo1,CondTITheo2}. Coupling of helical edge states with spin impurities or nuclear in topological insulator \cite{HiperCoupling,NuclearSpinTI} leads to the Anderson localization \cite{AndLoc} and anisotropic nuclear spin-spin interaction \cite{internalspin1DSHE} with spiral spin ordering \cite{NuclearSpinTI}. It is also demonstrated that Mn-doped InAs/GaSb \cite{Mn-doped} develops a ferromagnetic ordering at low enough temperature. Therefore, the effect of ferromagnetism generated by the magnetic impurities would gap out the boundary states spectrum \cite{Masive1,Masive2,Mn-doped,1DGapFormImpu}. Nevertheless, in spite of being 1D Dirac dispersion in some states of matter \cite{BHZTI,1DD} its properties has not been demonstrated fully with broken time-reversal symmetry in other aspects. It, thus, deserves to investigate indirect interaction between magnetic moments in magnetic topological insulators with Zeeman exchange splitting resulted from breaking of time-reversal symmetry due to spin polarization of topological edge states \cite{PlorizedTI}.

The purpose of this research is to study indirect exchange interactions between magnetic impurities mediated by 1D helical carriers of a two-dimensional topological insulator with weakly broken time-reversal symmetry. Here, we focus on the role of gapped spectrum of helical edge states. We find that unlike usual 1D quantum wire case \cite{1DRKKYConv}, the RKKY exchange interaction in our system is strongly anisotropic and includes the Heisenberg, Dzyaloshinsky-Moriya, and Ising terms which is somewhat similar to gapless spin-orbit-coupled quantum wire \cite{1DRKKYSpinOrbit} and gapful spin-orbit-coupled carbon nanotubes \cite{1DRKKYSpinOrbitCNT}. However, in contrast, here, the Ising terms are comprised of in-plane and out-of-plane interactions which have the same form as the case of three-dimensional topological insulators in the presence of bandgap \cite{3DTIRKKY0}. The corresponding range functions display an oscillatory (exponentially) decay with a spatial separation of magnetic moments if chemical potential resides in the band (bandgap). Furthermore, all the range functions are significantly affected by the energy gap, in particular, near the band edge leading to decreasing of oscillation period as a function of energy gap or Fermi energy. Moreover, in comparison with Heisenberg, Dzyaloshinsky-Moriya, and in-plane Ising terms having invariable oscillation amplitude independent of Fermi energy position with respect to energy gap, the oscillation amplitude of out-of-plane Ising term increases (decreases) as a function of energy gap (Fermi energy). For both subgap and over gap regimes analytical expressions are presented for asymptotic behaviors of the range functions. Also, electron-electron interactions are included.

The paper is organized as follows. The model and basic theory of indirect exchange interaction between magnetic impurities are introduced in Sec. \ref{s2}. In Sec. \ref{s3}, analytical and numerical results for non-interacting case are presented. The effects electron-electron interactions are analyzed in Sec. \ref{s4}. Section \ref{s5} is devoted to summarize.

\begin{figure}[!htb]
\begin{center}
\includegraphics[width=8.8cm]{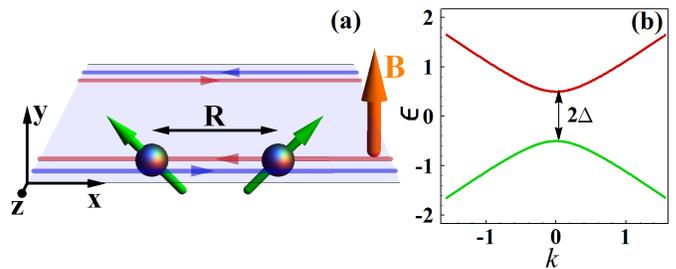}
\caption{(Color online) (a) Two-dimensional topological insulator with 1D edge states at two boundaries including two magnetic impurities on one of the edges separated by a distance $R$. The system is also influenced by Zeeman exchange field $B$. (b) Dispersion relation of the system edge states with gap 2$\Delta$.}
\label{2DTI}
\end{center}
\end{figure}

%%%%%%%%%%%%%%%%%%%%%%%%%%%%%%%%%%%%%%%%%%%%%%%%%%%%%%%%%%%%%%%%%%%%%%%%%%%
\section {Model and Theory}\label{s2}
%%%%%%%%%%%%%%%%%%%%%%%%%%%%%%%%%%%%%%%%%%%%%%%%%%%%%%%%%%%%%%%%%%%%%%%%%%%

The setup we consider throughout the paper is schematically depicted in Fig. \ref{2DTI}(a). It is composed of a two-dimensional topological insulator with 1D helical edge state and two magnetic impurities located on the edges of the topological insulator in the presence of Zeeman exchange field. Helical edge states of two-dimensional topological insulators can be thought of as 1D Dirac states. Also, inclusion of exchange field which is perpendicular to the surface gaps out the metallic edge states at the boundaries by violating time-reversal symmetry. Thus, we focus on the following Hamiltonian describing 1D gapped Dirac states \cite{1DGapFormImpu,HelicalZeemanHam1,HelicalZeemanHam2},
\begin{eqnarray}
H_{0}(k) &=& v_f k\sigma_x + \Delta \sigma_y,
\label{H0}
\end{eqnarray}
and the corresponding edge spectrum reads
\begin{eqnarray}
\varepsilon_{\lambda k} &=& \lambda \sqrt{(v_fk)^2 + \Delta^2},
\label{En}
\end{eqnarray}
where the Pauli matrices $\sigma_{x,y}$ represent the physical spin of the electron, $k$ is a wave vector along the 1D
channel, the band index $\lambda = \pm$, and $v_f$ is Fermi velocity. Since the spin quantization axis of edge states is along the x direction, the exchange field induced by out-of-plane spin polarization opens up gap $2\Delta$ in the edge spectrum (see Fig. \ref{2DTI}(b)). Thus, this lifts two-fold degeneracy of the so-called Dirac point. However, using magnetic insulator, such as EuS, EuO, EuSe, which is proximity coupled to the system can induce considerable Zeeman exchange field with gap size 9 meV \cite{ProxgapSize1, ProxgapSize2}. Also, It has been demonstrated that the fabricated MnBi2Se4/Bi2Se3 heterostructure exhibits ferromagnetism up to room temperature accompanied by a Dirac gap opening of $\sim$ 100 meV \cite{ProxgapSize3}. In contrast, the surface state gap varies between several tens to a hundred meV \cite{DopgapSize1} in magnetically doped topological insulators and its formation has been controversial depending on the impurity and impurity distance from the surface \cite{DopgapSize3}. However, the bandgap has been estimated to be as high as 10 meV Mn-doped InAs/GaSb \cite{Mn-doped,1DGapFormImpu} which can be suppressed by non-magnetic impurities \cite{1DGapFormImpu}. Note also that in topological insulators due to presence of strong spin-orbit interaction, the electron spin is no longer a good quantum number originating from the off-diagonal elements of BHZ Hamiltonian \cite{BHZTI}.

Two localized magnetic impurities with moments $\mathbf{S}_j (j = 1,2)$ can be coupled to itinerant spin-polarized Dirac fermions with spin density $\mathbf{s}(R_j) = \delta(r - R_j)\boldsymbol \sigma$ at position $R_j$. This coupling can be modeled by
\begin{equation}
H_{\text{int}} = J \sum_{j=1,2} \mathbf{S}_j \cdot \mathbf{s}(R_j),
\end{equation}
where $J$ denotes coupling strength of magnetic impurities with the host Dirac fermions. Using the second-order perturbation theory and treating $H_{\text{int}}$ as a perturbation, the RKKY interaction between two magnetic impurities mediated by host carriers can be obtained by \cite{RKKYPertu1,RKKYPertu2,RKKYPertu3}
\begin{eqnarray}
%\begin{split}
H_{\text{RKKY}} = &-&\frac{J^2}{\pi}Tr [ \int_{-\infty}^{\epsilon_f} d\epsilon\ Im\{(\mathbf{S_1}\cdot\boldsymbol{\sigma})\nonumber\\
&\times & G^0( R,\epsilon^+) (\mathbf{S_2}\cdot\boldsymbol{\sigma}) G^0(- R,\epsilon^+)\}],
\label{HRKKY}
%\end{split}
\end{eqnarray}
where $Tr$ stands for trace over spin degree of freedom, $Im$ is imaginary part, $\epsilon_f$ is the Fermi energy measured from the Dirac point. $G^0(R,\epsilon^+)$ denotes the Green's function matrix in real-space with $R = R_2 - R_1$ being a distance between the two magnetic centers and $\epsilon^+ = \epsilon+i \eta$ with $\eta\rightarrow0^+$. The real-space Green's function can be written as,
\begin{eqnarray}
G^0 (R, \epsilon^+)& = &\int \frac{dk}{2\pi} \ e^{i k R} G^0 (k, \epsilon^+),
\label{G-RE}
\end{eqnarray}
with
\begin{equation}
G^0(k, \epsilon^+) = \frac{\epsilon^+ + H_0(k)}{(\epsilon^+)^2-(v_fk)^2-(\Delta)^2},
\label{G-MO}
\end{equation}
where $G^0(k, \epsilon^+)$ is the momentum space Green's function associated with Eq. (\ref{H0}). Substituting Eq. (\ref{G-MO}) into Eq. (\ref{G-RE}) and using the residue theorem, we derive a closed form expression for the real-space Green's function as
\begin{eqnarray}
G^0 (\pm R, \epsilon^+)& = &-\frac{ie^{\frac{i\sqrt{\alpha}}{v_f}R}}{2v_f\sqrt{\alpha}}(\epsilon^+ \sigma_0 \pm \sqrt{\alpha} \sigma_x + \Delta \sigma_y),
\label{Green}
\end{eqnarray}
where $\alpha = (\epsilon^+)^2 - (\Delta)^2$ and $\sigma_0$ is the unit matrix. Combining the result in Eq. (\ref{Green}) with Eq. (\ref{HRKKY}), yields a formula for $H_{RKKY}$ as follows:
\begin{eqnarray}
H_{\text{RKKY}} =&F_1&(R,\epsilon_f) \mathbf S_1 \cdot \mathbf S_2 + F_2(R,\epsilon_f) (\mathbf S_1 \times \mathbf S_2)_x\nonumber\\
 &+& F_3(R,\epsilon_f) \mathbf{S}^x_1 \mathbf{S}^x_2 + F_4(R,\epsilon_f) \mathbf{S}^y_1 \mathbf{S}^y_2,
\label{finalHRKKY}
\end{eqnarray}
where the range functions are
\begin{eqnarray}
F_1 &=& Im \int_{-\infty}^{\epsilon_f} d\epsilon f(\alpha) ,\label{rangeF1}\\
F_2 &=& -Im \int_{-\infty}^{\epsilon_f} d\epsilon \frac{i \epsilon}{\sqrt{\alpha}} f(\alpha),\label{rangeF2}\\
F_3 &=& -F_1,\label{rangeF3}\\
F_4 &=& Im \int_{-\infty}^{\epsilon_f} d\epsilon \frac{\Delta(2i\sqrt{\alpha}+\Delta)}{4\alpha} f(\alpha),
\label{rangeF4}
\end{eqnarray}
with $f(\alpha) =\frac{J^2}{\pi v^2_f} e^{\frac{2i\sqrt{\alpha}}{v_f}R}$. As one can see from Eq. (\ref{finalHRKKY}), the RKKY interaction consists of the Heisenberg, the Dzyaloshinsky-Moriya, and two component Ising interactions whose range functions are given by $F_1$, $F_2$, $F_3$, and $F_4$, respectively. The Heisenberg, x- and y-component Ising interactions favor collinear magnetic spin alignment while the Dzyaloshinsky-Moriya imposes in-plane non-collinear magnetic spin orientation. Notably, the Heisenberg, Dzyaloshinsky-Moriya and x-component of Ising terms come from helical nature of the edge states and their range functions are modified due to gap opening [see Eqs. (\ref{rangeF1})-(\ref{rangeF3})]. In contrast to non-interacting \cite{2DTIRKKY1} and interacting \cite{NuclearSpinTI} gapless helical cases where out-of-plane component of the RKKY coupling is negligible compared to the in-plane ones, interestingly, Eq. (\ref{rangeF4}) implies that the out-of-plane component remains survived. This originates from partial mixing of opposite spin states of helical states due to gap term. It should also be noted that by setting $\Delta = 0$, we recover the previously obtained results \cite{2DTIRKKY1}. Furthermore, in the gap region, the integrands of the range functions become purely real functions, whereby all these terms will be suppressed, resulting actually from the lack of available states for itinerant carriers. However, if chemical potential lies within the gap region then, for sufficiently small energy gap, virtual interband transitions of electrons can mediate the indirect exchange interaction, which is known as Bloembergen-Rowland mechanism \cite{BRInt1,BRInt2}.

%%%%%%%%%%%%%%%%%%%%%%%%%%%%%%%%%%%%%%%%%%%%%%%%%%%%%%%%%%%%%%%%%%%%%%%%%%%
\section {Analytical and numerical results} \label{s3}
%%%%%%%%%%%%%%%%%%%%%%%%%%%%%%%%%%%%%%%%%%%%%%%%%%%%%%%%%%%%%%%%%%%%%%%%%%%

In what follows, due to particle-hole symmetry, without loss of generality, we assume that $\epsilon_f > 0$. In general, although there are no analytical expressions for the range functions [Eqs. (\ref{rangeF1})-(\ref{rangeF4})], it would be possible to estimate analytical statements for them in some limiting cases. In the small gap regime, {\it i.e.}, $\epsilon_f \gg \Delta$, Eqs. (\ref{rangeF1})-(\ref{rangeF4}) can be approximated by
\begin{eqnarray}
F_1 &\approx& -\frac{J^2}{\pi v_f}\left[\frac{1}{2 R}\cos(\gamma)+\frac{\Delta^2R}{v^2_f}Ci(\gamma)\right] ,\label{rangeF11}\\
F_2 &\approx& -\frac{J^2}{2\pi v_f} \left[\frac{1}{R}\sin(\gamma)-\frac{\Delta^2}{v^2_f\epsilon_F}\cos(\gamma)\right],\label{rangeF22}\\
F_3 &=& -F_1,\label{rangeF33}\\
F_4 &\approx& -\frac{J^2}{2\pi v^2_f}\Delta Ci(\gamma),
\label{rangeF44}
\end{eqnarray}
where $\gamma = \frac{2R\epsilon_f}{v_f}$ and $Ci(x)$ is cosine integral function \cite{table}. We note that the first term of range functions of the Heisenberg, Dzyaloshinsky-Moriya, and in-plane Ising interactions, having dominant contribution, decay as $R^{-1}$ reminiscing the range function behavior of usual 1D electron gas. Also, exploiting the asymptotic form of cosine integral function \cite{table},
\begin{eqnarray}
Ci(x)\approx \frac{1}{x}\left[\sin(x)-\frac{\cos(x)}{x}\right],
\label{Asemp}
\end{eqnarray}
for long distance limit, {\it i.e.}, $x,R \gg 1$, we identify that the spatial dependence of y-component of Ising interaction falls off similar to the other interactions.

\begin{figure}[t]
  \centering
  \includegraphics[width=8.5cm]{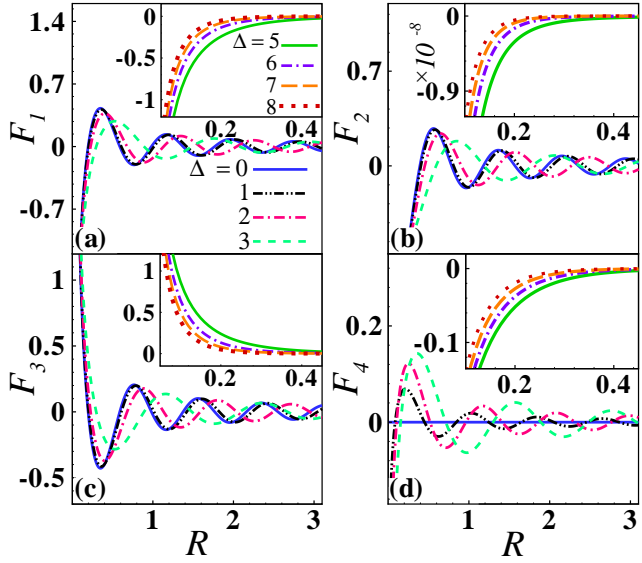}
  \caption{(Color online) Range functions versus $R$ for $\epsilon_f$ = 4 with different values of $\Delta$. Main panels are for $\epsilon_f > \Delta$ while for the insets $\epsilon_f < \Delta$.}
  \label{fig2}
\end{figure}

In the subgap limit, {\it i.e.}, $\epsilon_f \ll \Delta$, for long distance case, on the other hand, using steepest descent method \cite{SteepDec} we can determine leading-order asymptotic approximations to Eqs. (\ref{rangeF1})-(\ref{rangeF4}) as,
\begin{eqnarray}
F_1 &\approx&  -\frac{J^2}{2v_f^{\frac{3}{2}}} \sqrt{\frac{\Delta}{\pi R}} e^{-\frac{2\Delta R}{v_f}},\label{rangeF11A}\\
F_2 &\approx& -\frac{J^2\eta\epsilon_f}{\pi v_f^{2}\Delta} e^{-\frac{2\Delta R}{v_f}},\label{rangeF22A}\\
F_3 &=& -F_1,\label{rangeF33A}\\
F_4 &\approx& -\frac{J^2}{8v_f^{\frac{3}{2}}} \sqrt{\frac{\Delta}{\pi R}} e^{-\frac{2\Delta R}{v_f}}.\label{rangeF44A}
\label{AsempDbigE}
\end{eqnarray}
Note that, in this regime, the above obtained range functions have an exponentially decaying behavior determined by the energy gap and Fermi velocity.

\begin{figure}[t]
\begin{center}
\includegraphics[width=8.5cm]{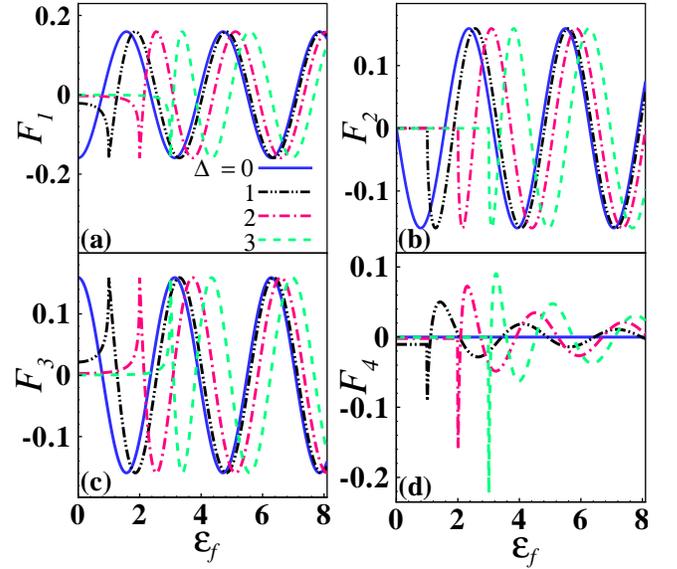}
\caption{(Color online) Range functions as a function of $\epsilon_f$ with different values of $\Delta$ for $R = 1$.}
\label{fig3}
\end{center}
\end{figure}

Generally, behaviors of the range functions can be explored by numerical evaluation. We set $v_f a^{-1}_B = 1$ by introducing Bohr radius $a_B$ and $J = 1$. The range functions F$_1$, F$_2$, F$_3$, and F$_4$ are plotted as a function of $R$ in Fig. \ref{fig2} for different values of $\Delta$ with $\epsilon_f = 4$. As shown in the main panels of Fig. \ref{fig2}, for $\epsilon_f > \Delta$ the oscillations of all the range functions are damped by increasing $R$ and the period of oscillations is increased with the increase of $\Delta$. In the meanwhile, the amplitude of all the range functions decreases with the increment of band gap, interestingly, except for the case of out-of-plane Ising term that increases. Moreover, for $\epsilon_f < \Delta$, as shown in the insets of Fig. \ref{fig2}, all the range functions decay exponentially with the distance of two magnetic impurities. Moreover, $F_2$ takes infinitesimally small values compared to the others. We also observe that the rate of exponential decay becomes faster with increasing $\Delta$ arising from suppression of the excitations of massive carriers across the gap. As a result, the interactions result in magnetic ordering of impurities depending on the values of the parameters. However, the spiral ordering due to Dzyaloshinsky-Moriya term is smeared out by the induced gap in the over gap regime. While in the subgap regime, the gap term completely suppresses the spiral ordering and, instead, establishes considerable out-of-plane magnetization. Since, as already mentioned above, the spiral ordering is originated from the helicity, this indicates that the gap opening gradually mixes the spin states and destroys the helicity.

The dependence of range functions on Fermi energy for various values of $\Delta$ with $R = 1$ is depicted in Fig. \ref{fig3}. For $\epsilon_f < \Delta$ the absolute values of range functions of both Heisenberg and in-plane Ising interactions gradually increase from small values with the enhancement of $\epsilon_f$ until reach a maximum at $\epsilon_f = \Delta$ as shown in Figs. \ref{fig3}(a) and \ref{fig3}(c). In the energies below $\Delta$, however, the magnitudes of Dzyaloshinsky-Moriya and out-of-plane Ising interactions are smaller than those of the other range functions [see Figs. \ref{fig3}(b) and \ref{fig3}(d)]. This results in dominating the ferromagnetic coupling in the subgap regime \cite{3DTIRKKY0}. Moreover, the out-of-plane Ising term shows a sharp dip at $\epsilon_f = \Delta$ which can be attributed to the high spin-polarized density of states available at the band edge. Moreover, all the range functions exhibit oscillatory behavior for $\epsilon_f > \Delta$ except that the envelope of $F_4$ is damped simultaneously. Also, $F_1$, $F_2$, and $F_3$ demonstrate constant amplitudes for different values of $\Delta$ while the amplitude of oscillations of $F_4$ increases as a function of $\Delta$.

\begin{figure}[t]
\begin{center}
\includegraphics[width=8.5cm]{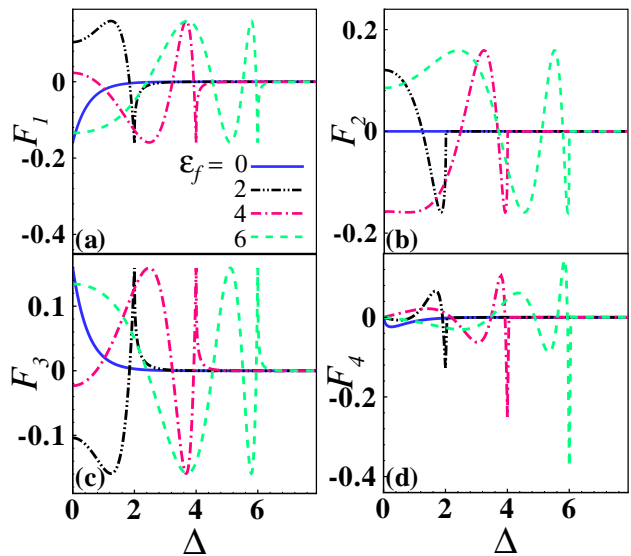}
\caption{(Color online) Dependence of range functions on $\Delta$ with different values of $\epsilon_f$ for $R = 1$.}
\label{fig4}
\end{center}
\end{figure}

The gap dependence of $F$'s for different values of $\epsilon_f$ is illustrated in Fig. \ref{fig4}. One can see that except for $F_4$ which is zero at $\Delta = 0$ [see Fig. \ref{fig4}(d)] the other range functions can take various values depending on both Fermi energy and $R$ [see Figs. \ref{fig4}(a), \ref{fig4}(b), and \ref{fig4}(c)]. By increasing $\Delta$ the values of $F$'s change with small modulations at first. Then, as $\Delta$ increases further the range functions begin to oscillate in $\Delta \lesssim \epsilon_f$. As already mentioned above, from Fig. \ref{fig4} it is also visible that the oscillation amplitude of y-component of Ising term is enhanced by increasing bandgap while the amplitude of oscillations of the other range functions remains unchanged even for different values of $\epsilon_f$. Note, interestingly, that the range function of Dzyaloshinsky-Moriya interaction is identically zero for $\epsilon_f = 0$ irrespective of $\Delta$, as illustrated in Fig. \ref{fig4}(b).

In both Figs. \ref{fig3} and \ref{fig4}, the period of oscillatory part of the range functions decreases by approaching to the band edge and, as a result, takes smallest values near band edge. When the Fermi energy is so small that $\Delta > \epsilon_f$ the range functions tend to zero rapidly. This is a consequence of vanishing of Fermi surface. Finally, it is worthwhile noting that the above-presented numerical behavior is in agreement with the analytical one quite well in the two limits of $\epsilon_f \gg \Delta$ and $\epsilon_f \ll \Delta$.

%%%%%%%%%%%%%%%%%%%%%%%%%%%%%%%%%%%%%%%%%%%%%%%%%%%%%%%%%%%%%%%%%%%%%%%%%%%
\section {Effects of electron-electron interaction} \label{s4}
%%%%%%%%%%%%%%%%%%%%%%%%%%%%%%%%%%%%%%%%%%%%%%%%%%%%%%%%%%%%%%%%%%%%%%%%%%%

The RKKY interaction in interacting quantum wires without \cite{RKKYeeWoutSO1,RKKYeeWoutSO2} and with \cite{RKKYeeWSO,RKKYeeWSO1} spin-orbit interaction has been studied before. These studies have also been extended to spin-orbit-coupled helical liquids \cite{RKKYeeHelicLiq}. Now, let us consider the effects of electron-electron interactions on RKKY interactions in the 1D massive helical states. Due to absence of Fermi points in the subgap regime we restrict ourself to the $\epsilon_f > \Delta$ case involving Tomonaga-Luttinger liquid model \cite{Giamarchi}. Close to band edge the gap $\Delta$ bends the dispersion, as can be seen from Fig. \ref{2DTI}(b), so it renormalizes the Fermi velocity as $v^{\prime}_f = v_f\sqrt{1-(\Delta/\epsilon_f)^2}$. We linearize \cite{BHZTI} the massive helical states of Eq. (\ref{H0}) on one of the system edges around Fermi points at momenta $\pm k_{f} =\pm \sqrt{\epsilon^2_f-\Delta^2}/v_f$ to obtain the low-energy Hamiltonian as
\begin{eqnarray}
\mathcal{H}_{0}
=-iv^{\prime}_f\left[\psi^{\dag}_{R}(x)\partial_x\psi_{R}(x)-\psi^{\dag}_{L}(x)\partial_x\psi_{L}(x)\right],
\label{H0TLL}
\end{eqnarray}
and the low-energy electron field operator as
\begin{eqnarray}
\psi(x) =\sum_r\chi_r(rk_{f})\psi_{r}(x)e^{irk_{f}x},
\label{psiTLL}
\end{eqnarray}
where the operators $\psi_{r}(x)$ with $r = R,L = +,-$ are the annihilation fields for right- and left-moving fermions at coordinate point $x$ along the edge. The spinor $\chi_{r}(k)$ in spin space is given by
\begin{eqnarray}
\chi_r(k) =\frac{1}{\sqrt{2}}\left(
           \begin{array}{c}
             re^{i\theta_k} \\
             e^{-i\theta_k} \\
           \end{array}
         \right),
\label{spinor}
\end{eqnarray}
with $\theta_k = 1/2 \tan^{-1}(-\Delta/kv_f)$ being $k$-dependent spin-rotation angle implying that the gap term mixes the spin states [Eq. (\ref{spinor})] so that the spin is no longer a good quantum number. The fermionic-field operators can be expressed in terms of bosonic fields $\phi_{r}(x)$ as \cite{Giamarchi}
\begin{eqnarray}
\psi_{r}(x) &=&\frac{1}{\sqrt{2\pi a_0}}\eta_{r}e^{i r\sqrt{4\pi}\phi_{r}(x)},
\label{Field}
\end{eqnarray}
where $a_0$ is a short-distance cutoff, and $\eta_{r}$ are Klein factors.

The Hamiltonian of electron-electron interactions having dominant contribution are given by
\begin{eqnarray}
\mathcal{H}_{e-e} &=& g_{2}\psi^{\dag}_{L}(x)\psi_{L}(x) \psi^{\dag}_{R}(x)\psi_{R}(x)\nonumber\\
&+&\!\frac{g_{4}}{2}\!\left[\left(\psi^{\dag}_{L}(x)\psi_{L}(x)\right)^2+\left(\psi^{\dag}_{R}(x)\psi_{R}(x)\right)^2\right],
\label{Hee}
\end{eqnarray}
where $g_{2,4}$ are interaction constants. We treat the total Hamiltonian $\mathcal{H}=\mathcal{H}_{0}+\mathcal{H}_{e-e}$ in the bosonized form yielding
\begin{eqnarray}
\mathcal{H} &=& \frac{v}{2}\left[\frac{1}{K}(\partial_x\Phi)^2+K(\partial_x\Theta)^2\right],
\label{HBOZ}
\end{eqnarray}
with bosonic fields
\begin{eqnarray}
\Phi = \phi_{R}+\phi_{L}, \quad
\Theta =\phi_{R}-\phi_{L}.
\end{eqnarray}
Here, we have introduced the speed of collective excitations $v$ and Luttinger parameter $K$ as
\begin{eqnarray}
v &=& \sqrt{(v^{\prime}_f+\frac{g_4}{2\pi})^2-(\frac{g_2}{2\pi})^2},\\
K &=& \sqrt{\frac{2\pi v^{\prime}_f+g_4-g_2}{2\pi v^{\prime}_f+g_4+g_2}},
\end{eqnarray}
depending on gap and Fermi energy through $v^{\prime}_f$ in contrast to the conventional ones. For noninteracting case $K = 1$ while $0 < K < 1$ ($K > 1$) corresponds to repulsive (attractive) electronic interactions. Also, we have assumed the weak interaction ignoring backscattering potential \cite{backscat1,backscat2}. Note, however, that it has been shown that elastic spin-flip backscattering can be occurred by coupling of electrons to nuclear spins through the hyperfine interaction causing a significant edge resistance at low enough temperatures \cite{NuclearSpinTI}.

In terms of electron field operator, the spin density can be defined as $\mathbf{s}(x)= \psi^{\dag}(x)\boldsymbol \sigma\psi(x)$. Using the identity
\begin{eqnarray}
\chi_r^{\dagger}(r k_{f})\boldsymbol \sigma \chi_{r^{\prime}}(r^{\prime} k_{f}) \!=\!\!\left(\!\!\!
                                                                    \begin{array}{c}
                                                                     r\delta_{r,r^{\prime}} \cos(2\theta_{k_f}\!) \\
                                                                      -ir\delta_{r,-r^{\prime}}\!+\!\delta_{r,r^{\prime}} \sin(2\theta_{k_f}\!) \\
                                                                       -\delta_{r,-r^{\prime}} \cos(2\theta_{k_f}\!) \\
                                                                    \end{array}
                                                                  \!\!\!\right),
\end{eqnarray}
and Eq. (\ref{Field}) the components of spin density $\mathbf{s}(x)$ can be determined in the bosonized language \cite{RKKYeeWoutSO1,RKKYeeWoutSO2} as
\begin{eqnarray}
s^x \!&=&\!\frac{1}{\sqrt{\pi}}\cos(2\theta_{k_f}\!)\partial_x\Theta,\\
s^y \!&=&\!\frac{1}{\sqrt{\pi}}\sin(2\theta_{k_f}\!)\partial_x\Phi\!+\!\frac{1}{\pi a_0}\cos(\sqrt{4\pi}\Phi\!+\!2k_fx),\\
s^z \!&=&\!-\frac{1}{\pi a_0}\cos(2\theta_{k_f}\!)\sin(\sqrt{4\pi}\Phi\!+\!2k_fx).
\end{eqnarray}
Having obtained the components of spin density and following the procedure in Ref. \cite{Giamarchi}, the imaginary-time spin-spin correlation functions
\begin{eqnarray}
\mathcal{S}^{ab}\!(\tau,R) = \langle s^a(\tau,R) s^b(0,0)\rangle,
\end{eqnarray}
can be determined yielding nonzero components as
\begin{eqnarray}
\mathcal{S}^{xx}\!(\tau,\!R)\!&=&\!\frac{\cos^2(2\theta_{k_f}\!)}{2\pi^2K}X\!(\tau,\!R),\\
\mathcal{S}^{yy}\!(\tau,\!R)\!&=&\!\frac{\sin^2(2\theta_{k_f}\!)K}{2\pi^2}\!X\!(\tau,\!R)\!+\!\frac{\cos(2k_f\!R)}{2\pi^2a^2_0}\!Y\!(\tau,\!R)\!,\\
\mathcal{S}^{zz}\!(\tau,\!R)\!&=&\!\frac{\cos^2(2\theta_{k_f}\!)\!\cos(2k_fR)}{2\pi^2a^2_0}Y(\tau,\!R),\\
\mathcal{S}^{yz}\!(\tau,\!R)\!&=&\!-\mathcal{S}^{zy}\!(\tau,\!R)\!=\!\frac{\cos(2\theta_{k_f}\!)\!\sin(2k_f\!R)}{2\pi^2a^2_0}\!Y\!(\tau,\!R)\!,
\end{eqnarray}
with
\begin{eqnarray}
X\!(\tau,\!R)&=&\left[\frac{(v|\tau|+a_0)^2-R^2}{[(v|\tau|+a_0)^2+R^2]^2}\right],\\
Y\!(\tau,\!R)&=&\left[\frac{a_0^2}{(v|\tau|+a_0)^2+R^2}\right]^K.
\end{eqnarray}

By integrating $\mathcal{S}^{ab}\!(\tau,R)$ over $\tau$, i.e.,
\begin{eqnarray}
\kappa^{ab}(R)=\int^{\infty}_{-\infty} d\tau\mathcal{S}^{ab}(\tau,\!R),
\end{eqnarray}
one gets spin susceptibility. Plugging $\kappa^{ab}(R)$ into the following general formula for RKKY interaction
\begin{eqnarray}
H_{RKKY}=-J^2\!\!\!\!\!\!\sum_{a,b=x,y,z}\!\!\!\!\!\!\kappa^{ab}(R)S_1^aS_2^b,
\end{eqnarray}
we obtain the similar magnetic spin structure as Eq. (\ref{finalHRKKY}) with modified range functions which for $a_0\ll|x|\ll v_f/k_BT$  given by
\begin{eqnarray}
F_1 &=&  -\frac{J^2\nu(K)}{2\pi R^{2K-1}} \cos^2(2\theta_{k_f}\!)\cos(2k_fR) ,\label{rangeF11AI}\\
F_2 &=& -\frac{J^2\nu(K)}{2\pi R^{2K-1}} \cos(2\theta_{k_f}\!)\sin(2k_fR),\label{rangeF22AI}\\
F_3 &=& -F_1,\label{rangeF33AI}\\
F_4 &=& -\frac{J^2\nu(K)}{2\pi R^{2K-1}} \sin^2(2\theta_{k_f}\!)\cos(2k_fR),\label{rangeF44AI}
\label{AsempEbigD}
\end{eqnarray}
where $\nu(K)=\frac{a^{2K-2}_0\Gamma(K-1/2)}{\sqrt{\pi}v\Gamma(K)}$ with $\Gamma(p)$ being Gamma function. Consequently, we observed that since the spin basis of Hamiltonian (\ref{H0TLL}) is the same as Hamiltonian (\ref{H0}) so the spin texture structure of Eq. (\ref{finalHRKKY}) remains unchanged for interacting electrons in the host Luttinger liquid and only the range functions are modified. This modification affects on the power-law decay characterizing by exponent $2K-1$ and on the magnitude of range functions. Note that the above equations are valid in the range 1/2 $<$ K $\leq$ 1 below which RKKY-Kondo competition becomes important \cite{RKKYKondeeHelicLiq}. Moreover, for non-interacting case, i.e., $K = 1$, and $\epsilon_f \gg \Delta$ expanding Eqs. (\ref{rangeF11AI})-(\ref{rangeF44AI}) provides similar leading behavior as Eqs. (\ref{rangeF11})-(\ref{rangeF44}).

%%%%%%%%%%%%%%%%%%%%%%%%%%%%%%%%%%%%%%%%%%%%%%%%%%%%%%%%%%%%%%%%%%%%%%%%%%%
\section {Summary} \label{s5}
%%%%%%%%%%%%%%%%%%%%%%%%%%%%%%%%%%%%%%%%%%%%%%%%%%%%%%%%%%%%%%%%%%%%%%%%%%%

In summary, the RKKY interaction of interacting spins is investigated in the edge of time-reversal broken two-dimensional edge states. The resulting interaction is composed of Heisenberg, Dzyaloshinsky-Moriya, in-plane and out-of-plane Ising terms. The corresponding range functions fall off with a power-law decay $R^{-1}$ and exponentially in the over gap and subgap regimes, respectively, at large distance. As the energy gap increases, the rate of spatial exponential decay and the spatial periodicity of the range functions increase. The range functions of these interactions are significantly influenced through gapped band structure providing small periods near the band edge as a function of Fermi energy or energy gap. We found if the Fermi energy lies inside the bandgap then the range function of Dzyaloshinsky-Moriya interaction has much smaller values than those of other interactions. Also, the amplitude of range function of out-of-plane Ising interaction increases as gap increases for finite Fermi surface while the amplitudes of the other interactions do not change. In the some limiting cases analytical results are approximately extracted for the range functions supporting the numerical results. With taking into account electron-electron interactions it is shown the magnitude and power-low decay of the range functions change depending on collective excitations velocity and Luttinger parameter. It is worthwhile noting that by combining density functional theory calculations with complementary experimental techniques, it has been revealed that transition metal dopants not only affect the magnetic state of host material, but also alter the electronic structure \cite{DopgapSize3}. So, we believe that the ferromagnetic ordering of impurities which is dominated in the subgap regime supports the bandgap. Thus the renormalized spectrum will impact the interactions between impurities promoting ferromagnetic ordering as long as the Fermi surface resides within the bandgap. Finally, we remark that our setup can be implemented by source-probe measurement method \cite{source-probe} for low dimensional systems and be served as an alternative approach in investigating magnetic properties of edge states.

\end{document}